\documentclass[twocolumn]{revtex4}
\usepackage[T1]{fontenc}
\usepackage[latin1]{inputenc}
\usepackage{graphicx}

\begin{document}

\title[Surface Mount]{Reflow Soldering of Surface Mount Electronic Components in a Laboratory}

\author{Christopher J. Erickson, Dallin S. Durfee}

\affiliation{Department of Physics and Astronomy \\ Brigham Young University, Provo, UT 84602}

\email{erickson.cj@gmail.com, dallin_durfee@byu.edu}

\begin{abstract}

We present a basic tutorial for implementing surface mount technology in lab-built scientific instruments. We discuss the advantages and disadvantages of using surface mount chips. We also describe methods for the development and prototyping of surface mount circuitry in home-built electronics. The method of soldering surface mount components in a common toaster oven is described. We provide advice from our own experience in developing this technology, and argue that surface mount technology is often preferable to using leaded components when building large circuits, and is  essential if the desired component characteristics are only available in surface mount packages.

\end{abstract}

\maketitle

\section{Introduction}
Thanks to the consumer markets of devices such as cellular phones, GPS receivers, and hand-held PDAs, integrated chip makers have developed many high-speed, high-bandwidth, low-noise, and low-power consumption chips to fuel new technology. The characteristics of some of these chips make them ideal for use in sensitive scientific instruments. While scientific laboratories stand to benefit from the capabilities of the new technology, many of these chips are only available in surface mount packages designed for use in an automated assembly process.

Including surface mount components in circuits appears troublesome because the standard laboratory method of soldering electronic components by hand is messy and extremely time consuming for surface mount chips on a printed circuit board (PCB). Surface mount parts tend to be small with closely spaced pins, and some components have pin pads in the center of the chip that are completely inaccessible with a soldering iron. Furthermore, these chips must be soldered directly to the board without the use of sockets, which increases the risk of damage. This also increases the time, effort, and frustration that go into debugging a prototype where parts may need to be removed or changed out several times.  Purchasing industry standard equipment to perform the soldering is expensive. Table top wave-flow soldering stations start at a few thousands of dollars and are designed for mass production \cite{manncorp}, an unnecessary capability for most laboratories.

We set out to find an easy and inexpensive way to implement surface mount technology in our designs. What we found was that lab-built electronic devices based on surface mount chips are not only possible, but in many cases preferable. Obviously, surface mount electronics are necessary when required parts are only available in surface mount packaging. However, a surface mount package's low profile and short leads help reduce parasitics, such as stray inductance and capacitance, as well as coupling to ambient noise. Surface mount boards are also preferable to through-hole designs when making more than one copy of a circuit with many components. We present useful techniques for prototyping and debugging surface mount electronics, and two inexpensive methods for reflow soldering surface mount components on a PCB, one of which is already in use among some electronics hobbyists \cite{Maxon}.

\section{Advantages and Disadvantages}

There are several barriers to overcome when implementing surface mount chips into laboratory circuits, the most imposing of which is prototyping. Generally, electronics are developed on a breadboard where passive components and integrated circuits (ICs) can be easily interchanged without the use of a soldering iron. This makes it easy to optimize and finalize the design for the electronic circuit before building a more permanent version on a copper-clad board. In contrast, surface mount chips do not slip in and out of sockets and each version of the circuit must be soldered to a custom PCB. Additionally, surface mount pads are prone to separating from the PCB when chips are repeatedly soldered to and removed from them.

Surface mount parts are typically much smaller than leaded components, making them harder to handle and making it more difficult to read part numbers and codes. In particular, surface mount capacitors do not have values printed on them and many of them are similar in appearance. For these reasons, if a specific surface mount component is not required and the circuit contains only a few components, a surface mount design is not desirable.

Conversely, using surface mount chips can be much more efficient when assembling larger circuits. Many chips can be placed on the board and then soldered all at one time, as opposed to soldering each individual component one at a time. Their smaller size allows for more compact circuitry and a more efficient use of space. Also, unlike through-hole components, every chip in a surface mount circuit only contacts one side of the PCB. This not only makes it easier to remove a chip, but allows a chip to be removed without removing the PCB from its housing (see section \ref{sec:debug}).

In short, a circuit design with few components or that is a one time build probably gains little from implementing a surface mount design unless a necessary chip is only available in a surface mount package. However, circuits with many components, and especially standard designs for commonly used circuits, are perfect candidates for employing surface mount technology.

\section{Circuit Design and Layout}

There are several methods that we use to streamline the design process, which we discuss below. We also discuss circuit design and layout techniques that have simplified the debugging and optimization processes.

\subsection{SPICE Modeling}

Because it takes significant time and cost to manufacture a PCB, we recommend characterizing the circuit with a SPICE model first. SPICE is a standard numerical modeler for circuits and several versions are available for free from chip manufacturers. These typically come with extensive libraries of the chips made by that company, as well as the ability to load SPICE models for chips from other makers. SPICE modeling can be useful for quickly checking different  combinations of chips and other components to optimize the circuit. Also, when debugging a circuit, SPICE can be used to model the effect of probes on the circuit's performance.

\subsection{Prototyping with Leaded Components}

If through-hole components are available, the design can still be prototyped on a springboard using leaded components. Prototyping in this way allows for quick adjustment of the physical circuit design before manufacturing a PCB. However, this method may not work well for highly specialized or high-speed electronics where path lengths and feedback loops are critical.
 
\begin{figure}
    \begin{center}
        \includegraphics[width=.4\textwidth]{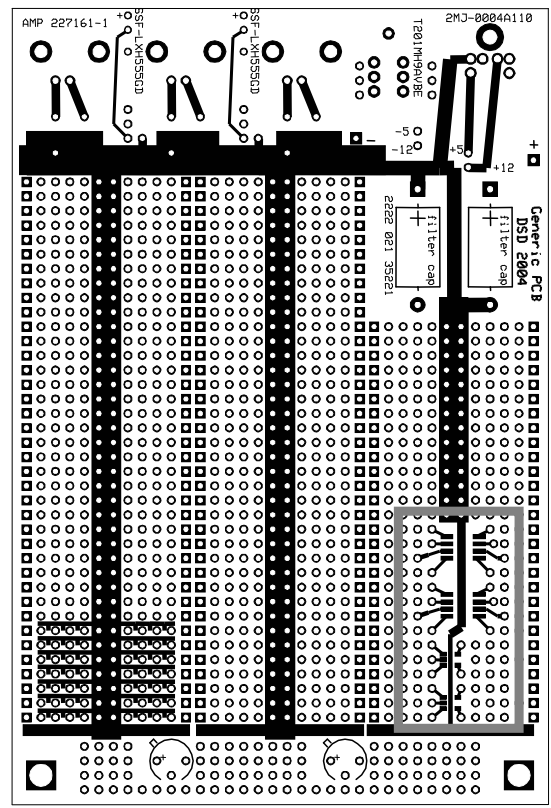}
    \end{center}
    \caption{Layout of one side of our prototyping board that allows for surface mount components to be integrated into the prototyping process. The surface mount section of the board is outlined by the gray box. \label{cap:proto}}
\end{figure}

We have also taken the time to design our own copper clad prototyping board with a section on it for surface mount components with various pin configurations (shown in figure \ref{cap:proto}). With these boards we can take advantage of sockets and leaded passive components to prototype the majority of the circuit while still using key surface mount chips.

\subsection{ExpressPCB}

Several resources exist for the design of the circuit, its PCB layout, and its production. ExpressPCB (\url{http://expresspcb.com}) offers two free CAD programs to aid in board design and production. The first, ExpressSCH, is a program for drawing the circuit schematics that also checks for inconsistencies such as broken wires. This program comes with a library of common electronic components and allows for the creation of new components. 

The schematic file can then be linked to a second program for design of the PCB layout, ExpressPCB, to help ensure the physical circuit is wired correctly. ExpressPCB comes with a library of pad layouts for various standard packages as well as specific chips accompanied by their Digikey part number. It also allows for the creation of new components. The program offers such features as highlighting pins that should be connected together and providing warnings for missing components. It will also print the board in its actual dimensions. Unfortunately, the program does not output an industry standard Gerber file so boards can only be ordered through ExpressPCB.

\subsection{Common Part Sizes That We Use}

Many surface mount components come in a variety packages. For op-amps and other active components, small outline integrated circuit (SOIC) packages with 8 or 14 pins and 0.05 inch pin spacing (pitch) are common and are easier to deal with than packages with smaller pitch. For passive components we found 1206 and 0805 chip sizes were common and not too small to handle. We have reserved 0805 chip sizes for precision components and 1206 sizes for standard components in our lab to avoid confusion when stuffing boards.

\subsection{Spacing between components}

While surface mount components are small and can be packed tightly together, it is good practice to leave a generous amount of space between the pads when designing a board. This prevents confusion as to which two pads define a passive component when stuffing the board, and it also makes it easier to correct errors that occur during the soldering process (see section \ref{sec:build}). Care should also be taken to allow adequate room around through-hole components, which can overshadow or cover surface mount parts on a poorly designed board, making it harder to access them if needed.

\subsection{Physical Design of the PCB\label{sec:layout}}

A well thought-out PCB design can simplify the debugging process. When designing a PCB it is generally recommended that traces run the width of the board on one side, and the length of the board on the other. This helps to keep the design simple and avoids the need to weave traces around each other with vias.

Place a myriad of ground vias on the board, and especially around sensitive components. This will introduce many small, closely-placed ground loops, which negate large ground loops and minimize inductance between the components and ground. Also, clearing ground planes from under fast chips and around traces carrying sensitive signals will reduce problems created by stray capacitance. Checking over the board to eliminate islands (areas of metal that are not connected to either ground or a trace) is also a good idea.

There are also a few things that can be done when designing the PCB to make it easier to stuff. It helps to lay out the circuit in an intuitive manner so that different areas of the board are easily recognizable by function. Also, we label all components with either their part number or their value (for passive chips), and we make short notes or legends on a blank area of the board to aid in assembling the circuit. This can be done with a silk screening process, which adds extra cost to the production, or by simply writing such information into the copper ground plane. Also, as with standard circuit design, active components should all be placed so that pin 1 is in the same orientation for every IC. Place as many surface mount components on the same side of the board as possible. Only one side of the board can be soldered at once and any components on the opposite side will have to be soldered by hand.

\begin{figure}
    \begin{center}
        \includegraphics[width=.4\textwidth]{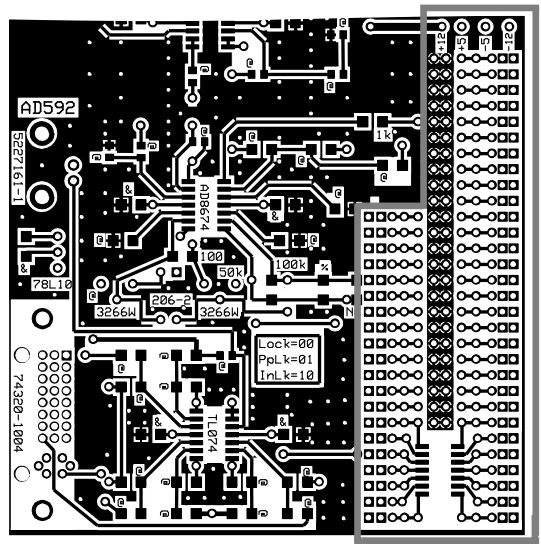}
    \end{center}
    \caption{The gray box encloses a prototyping grid that was added to a temperature controller circuit. This area can be used for debugging and/or adding components to the board including an SOIC chip. Power for this area can be accessed from the vias in the upper right corner of the figure, which provide $\pm$5 and $\pm$12 volts. \label{cap:debugarea}}
\end{figure}

Finally, we recommend including a small prototyping area on the PCB where ground, power, and extra pin pads can support additional components that might be needed after the initial implementation. An example of one is shown in figure \ref{cap:debugarea}.

\section{Building the Circuit}

\subsection{Layout of Surface Mount Components \label{sec:build}}

When building surface mount circuitry, all of the chips on one side of the PCB can be set in place at the same time with solder paste before being placed in an oven to complete the soldering process. Solder paste is a semi-solid mixture composed mainly of solder particles and flux. It is solid enough to hold its form, but soft enough to hold chips in place when they are pressed into it on a PCB. We found it easiest to use a 20 gauge tip on a syringe to apply the paste to individual pads. A component can then be carefully placed on the board using a pair of tweezers. 

Taking care when placing components on the board can reduce errors when the solder is melted. The majority of such problems are the result of either too much solder paste or an uneven amount of solder paste being placed on the different pins of a component. Too much solder paste will form solder bridges to adjacent pads or the ground plane during the cooking process. Placing more solder one side of the chip than the other can cause chips to rotate, slide, or tombstone (lift one side of the chip off of the board). 

For components with a pitch less than .05 inch, it may be easiest to place a bead of solder across all the pins. This will result in solder bridges, but solder bridges can be corrected after the cooking process (see section \ref{sec:debug}).

Typically, it takes less than an hour to stuff a PCB containing on the order of a 100 surface mount components once a person has a little practice.

\subsection{Through-Hole and Additional Surface Mount Components}

Some circuits will contain through-hole components and may even have surface mount chips placed on both sides of the board. We layout the side of the board with the most surface mount components on it first and solder those pieces in the toaster oven. Afterward, we will add through-hole components and the remaining surface mount parts by hand.

\subsection{Reflow Soldering}

We experimented with two methods of melting the solder paste: using a common toaster oven, and using a hot plate and heat gun. We perform all of our surface mount soldering using the toaster oven method, but also took data with the hot plate method for comparison. 

\subsubsection{Soldering in a Toaster Oven}

As far as we are aware, the idea of using a common toaster oven for surface mount soldering was first put forth in a newsletter of the Seattle Robotics Society \cite{Maxon}. The group insisted that this method is only made possible by the use of an expensive, water-soluble solder paste from Kester that requires refrigeration and has a short shelf-life. We found that while the water-soluble paste did work, its storage and shelf life were inconvenient and its flux left a slightly sticky residue on the PCB. In searching for an alternative, we found that ChipQuik's 63/37 w/flux solder paste worked just as well. The ChipQuik paste features a "no-clean" flux that leaves no residue on the board. Additionally, it is cheap and requires no refrigeration. While the official shelf-life of the paste is 6 months without refrigeration, we were able to use an unrefrigerated syringe of the paste over a year after its original use without any problems.

According to \cite{Maxon}, after placing a prepared board in the cold oven, the oven's temperature needs to be increased discretely in carefully timed intervals. This is done to allow the oven to come to equilibrium and, presumably, to prevent overshoot at each step. We investigated whether this was necessary or whether setting the oven to the maximum temperature setpoint from the start would work just as well.

The temperature across a blank, previously unused PCB was measured by strapping five K-type thermocouples across it. The thermocouples were connected to an Agilent 34970A Data Aquisition Switch Unit, which collected the data during the process. We then heated the board using an Oster model 6232-015 1500 Watt toaster oven.

\begin{figure}
    \begin{center}
        \includegraphics{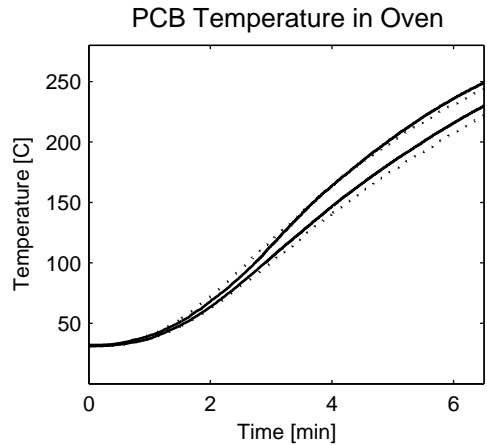}
    \end{center}
    \caption{Five thermocouples measured the temperature across a blank PCB while it was run through the soldering process using two different methods. The solid lines represent the maximum and minimum of the five temperature points observed on the board while the setpoint was changed in timed steps: 250~$^{\circ}$F (121~$^{\circ}$C) for 4 minutes, 375~$^{\circ}$F (191~$^{\circ}$C) for 2 minutes, and 450~$^{\circ}$F (232~$^{\circ}$C) for 30+ seconds to melt and bead the solder. The dotted lines represent the maximum and minimum profiles observed while heating the board with the oven temperature set to 450~$^{\circ}$F (232~$^{\circ}$C) from the start.\label{cap:gradual}}
\end{figure}

The maximum and minimum temperature data from five points across the PCB during two trials is shown in figure \ref{cap:gradual}. Even though the temperature was changed in steps at the intervals given in \cite{Maxon} for one data set, there is no evidence of it in the figure, indicating that the temperature setpoint is changed before the oven approaches equilibrium at any step. No gradual steps are needed in cooking the board, all that is required is to get the solder to its melting point, after which the board can be cooled. The entire cooking process takes about 6$\frac{1}{2}$ minutes per board with our oven.

The most temperature sensitive chip we use is an AD5541 16-bit digital to analog converter. Its maximum lead temperature of 215~$^{\circ}$C for 60 seconds \cite{AD5541} seems to be exceeded by the 246~$^{\circ}$C oven temperature seen when we took the data shown in figure \ref{cap:gradual} with no chips on the PCB. However, we have used the toaster oven to successfully create the low-noise, high-modulation bandwidth current drivers described in \cite{Erickson08}, as well as low-noise, high-bandwidth PID lock-circuits, high-speed homodyne photodetectors, and microprocessor control units. We have repeated this process multiple times and have never damaged a sensitive integrated chip through the process. Given that no chips have been damaged, we can infer that lead and chip temperatures are kept below their maximum limit. This may be due to the thermal mass of the chips and the evaporation of the liquid components of the solder paste.

When soldering a PCB with this method it is important to make sure that the rack inside the oven is level. This prevents gravity from moving any of the chips off of their pads once the solder melts. The PCB with its components and solder paste is placed in the oven at room temperature and the oven is turned on and set to 450~$^{\circ}$F (232~$^{\circ}$C). During the process of cooking the board, the solder paste flattens out as the oven begins to heat. Near the 5$\frac{1}{2}$ minute mark for our oven, solder will begin to melt, bead around individual pins, and become shiny. Once solder has beaded on all the pins, the oven can be turned off and the door opened. While \cite{Maxon} advises tapping the board immediately after cooking, we found that bumping, tipping, or jarring the board and oven immediately after the cooking process can cause chips to slide, rotate, and solder to adjacent pads, traces, or ground planes.  Allowing a few seconds before actually moving the board will allow the solder to cool enough so that chips do not reposition themselves.

\subsubsection{Soldering with a Hot Plate and Heat Gun}

We tested a second method of doing surface mount soldering in the lab by bringing the PCB up to a uniform temperature and then locally heating the PCB to melt the solder paste. A commercial system for this approach is available and includes a heated board rack and a hot air pencil for melting the solder at individual pins \cite{chipquik}. We tested to see if this method could be successfully implemented using a heat gun and hot plate already available in our lab. First, we prepared the PCB as described for the toaster oven and then heated it to $\sim$150~$^{\circ}$C on a hot plate. This is just below the 183~$^{\circ}$C melting point of the solder. By then passing a standard heat gun across the board the solder was brought to its melting point. 

\begin{figure}
    \begin{center}
        \includegraphics{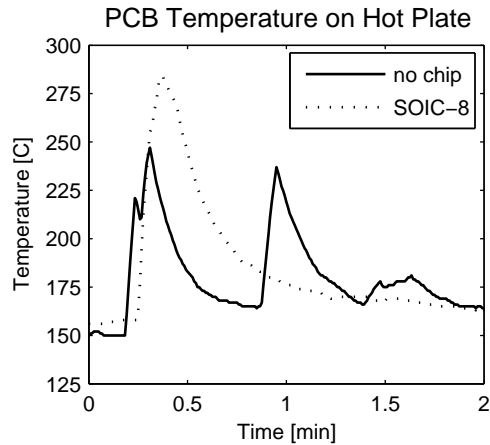}
    \end{center}
    \caption{After the PCB is brought to a 150 $^{\circ}$C, solder is melted by a heat gun to both the grounding plane and a trace with and without components present. The solid line shows temperatures spikes that occur from melting solder to a blank area of the board, while the dotted line shows the temperature spike due to soldering an 8 pin SOIC op-amp. Although the temperature spike for soldering the chip is short and local, the board and chip reach temperatures potentially dangerous to the chip. \label{cap:HPnochip}}
\end{figure}

We took data using a blank PCB with five thermocouples as in the oven trials. Figure \ref{cap:HPnochip} shows data taken by melting solder paste with a heat gun after the board had been brought to $\sim$150 $^{\circ}$C with a hot plate. The solid trace shows local temperature spikes that occur while melting solder to a blank area on the board, while the dotted trace shows the temperature spike from actually soldering an 8 pin SOIC package with this method.

We found that when using non-specialized equipment this method is time consuming and potentially dangerous to chips. The process of getting the hot plate to the right temperature with its crude dial took about 20 minutes, while actually soldering a single chip took 10 to 20 seconds. Overall, it would take the better part of an hour to solder one board. Also, the broad nozzle of the heat gun prevented accurate targeting of chip leads, which led to overheating of the entire chip. We have never used this method for actually building surface mount circuitry.

\section{Moving or Replacing Components \label{sec:debug}}

When the board has cooled, solder bridges, loose pins, tombstones, and misplaced pieces can be checked for. The use of a magnifying glass and a bright light are invaluable here. Usually, errors such as a solder bridges can be corrected by simply touching a soldering iron to the spot and dragging the solder onto the traces. However, some errors require removing solder from the board or resetting a chip. Also, some chips may need to be changed out if damaged through a design error or by later abuse in the circuit's life. Components with only two leads, such as capacitors and resistors, are often easily removed by gently tugging on them with a pair of fine tweezers while heating first one end, then the other with a soldering iron. Additional solder can be added to one of the leads to increase its heat capacity so that it stays molten longer. 

Removing components with more than two pins can be done through one of two methods. The first method is described in \cite{smttutor}. A braided copper desoldering wire or a desoldering pump is used to remove excess solder, as is commonly done on circuit boards of all designs. Then, a thin strand of enameled or nichrome wire can be threaded around the pins. Gently tugging on the wire while heating a pin will cause the pin to come off of its pad. This method is time consuming and can easily result in a damaged pad or pin.

A better way to desolder chips is to use ChipQuik Removal Alloy \cite{chipquik}. This is an easy and clean way to remove most surface mount chips. The alloy mixes with the solder and creates a low melting point eutectic that will stay liquid for some time after having been heated by a soldering iron. By applying a small amount of the removal alloy to all the pins of a component, the component can simply be lifted off of the board. The alloy can then be removed from the board with a desoldering pump or wire braid. Care should be taken when using the desoldering pump since its spring action combined with heat from the iron can easily lift or rip pads and traces off of a PCB.

Neither of these methods works well for chips with pads on the under side of the chip. If possible, we recommend avoiding such surface mount packages for this reason.

\section{Conclusion}

Using a toaster oven to solder surface mount PCB electronics is fast and safe. We have produced dozens of circuits with this method and it has consistently produced beautiful circuits with no cold solder joints and no dead chips.

We have found that implementing surface mount technology in lab-built devices is not only viable, but often preferable. Prototyping for these types of circuits can take longer than for through-hole circuits. However, once an electronic device is developed, the time and cost required for assembling and building a surface mount based device are lower.

Surface mount technology opens up a whole new world of electronic solutions for scientific instruments. Using this technology in the laboratory to update existing electronic devices and develop new ones can allow for better process control, more precise measurement, higher speed, and lower noise in experiments.

We would like to acknowledge the assistance of Brian Neyenhuis, Greg Doermann, and Ryan Dalrymple. This work was funded in part by the National Institute for Standards and Technology and the Air Force Research Laboratory.

\end{document}